\documentclass[12pt]{amsart}

\theoremstyle{plain}

\theoremstyle{definition}
\theoremstyle{definition}

\newcommand{\be}{\begin{equation}}
\newcommand{\ee}{\end{equation}}
\newcommand{\bea}{\begin{eqnarray}}
\newcommand{\eea}{\end{eqnarray}}
\newcommand{\bean}{\begin{eqnarray}}
\newcommand{\eean}{\end{eqnarray}}
\newcommand{\ba}{\begin{array}}
\newcommand{\ea}{\end{array}}
\newcommand{\ep}{\epsilon}

\newcommand{\Ga}{\Gamma}
\newcommand{\ga}{\gamma}

\newcommand{\la}{\lambda}

\newcommand{\pa}{\partial}

\newcommand{\no}{\nonumber}
\newcommand{\eq}{\eqref}
\begin{document}

 \title[waterbag model]
{On the water-bag model of dispersionless KP hierarchy}

 \author[Jen-Hsu Chang]
 {Jen-Hsu Chang  \\
 Department of Computer Science\\
 National Defense University\\
 Tauyuan, Taiwan\\
 E-mail: jhchang@ccit.edu.tw}
\maketitle
\begin{abstract}
We investigate the bi-Hamiltonian structure of the waterbag model of dKP
for two component case. One can establish the third-order and first-order  Hamiltonian
operator associated with the waterbag model. Also, the  dispersive corrections are discussed.
\end{abstract}
\section{Introduction}
The dispersionless KP hierarchy(dKP or Benney moment chain) is  defined by
\begin{equation}
\pa_{t_n}\la(p)=\{\la(p), B_n(p)\},\quad n=1,2, \cdots., \label{kpp}
\end{equation}
where the Lax operator $\la(p)$ is
\begin{equation}
\la(p)=p+\sum_1^{\infty} v_{n+1} p^{-n}  \label{pl}
\end{equation}
and it can be used to define a set of polynomials:
\[B_n(p)=\frac{[\la^n(p)]_{+}}{n},\quad i=1,2,3, \cdots  \quad t_1=x. \]
Here $[\cdots]_{+}$  denotes a non-negative part of the Laurent series $\la^n(p)$.
For example,
\[B_2=\frac{p^2}{2}+v_2,\qquad B_3=\frac{p^3}{3}+v_2 p+v_3.\]
Moreover,
the bracket  in  \eqref{kpp} stands for the natural Poisson bracket on the space of functions of
the two variables $(x,p)$:
\begin{equation}
\{f(x,p), g(x,p)\}=\pa_x f \pa_p g-\pa_x g \pa_p f. \label{po}
\end{equation}
The compatibility of  \eqref{kpp}  will  imply the zero-curvature equation
\begin{equation}
\pa_m B_n (p)-\pa_n B_m (p)=\{B_n (p),B_m(p)\}. \label{ze}
\end{equation}
If we denote $t_2=y$ and $t_3=t$, then the equation \eqref{ze} for $m=2, n=3$ gives
\begin{eqnarray*}
v_{3x}&=&v_{2y} \\
v_{3y}&=& v_{2t}-v_2v_{2x},
\end{eqnarray*}
from which the dKP equation is derived($v_2=v$)
\begin{equation}
v_{yy}=(v_t-vv_x)_x.\label{dkp}
\end{equation}
According to the  dKP theory \cite{Ak,KG, kr2, Ta1}, there exists a wave function $S(\la,x,t_2,t_3, \cdots)$ such that $p=S_x$ and satisfies
the Hamiltonian-Jacobian equation
\begin{equation}
\frac{\pa S}{\pa t_n}=B_n(p)\vert_{p=S_x}. \label{hj}
\end{equation}
It can be seen that the compatibility of \eq{hj} also  implies the zero-curvature equation \eqref{ze}.
 Now, we expand $B_n(p)$ as
\[B_n(p(\lambda))=\frac{[\la^n(p)]_+}{n}=\frac{\la^n}{n}-\sum_{i=0}^{\infty}G_{in}\la^{-i-1},\]
where the coefficients can be calculated by the residue form:
\[G_{in}=-res_{\la=\infty}(\la^i B_{n}(p)d\la)=\frac{1}{i+1} res_{p=\infty}(\la^{i+1} \frac{\pa B_n(p)}{\pa p}dp),\]
which also shows the symmetry property
\[G_{in}=G_{ni}.\]
Moreover, from
\[\frac{\pa B_m(\la)}{\pa t_n}=\frac{\pa B_n(\la)}{\pa t_m}\]
this implies the integrability of $G_{in}$ as expressed in terms of the free energy $\mathcal{F}$ (dispersionless $\tau$ function)\cite{Ta1}
\[G_{in}=\frac{\pa^2 \mathcal{F}}{\pa t_i \pa t_n}.\]
For example, the series inverse to \eq{pl} is
\begin{equation}
p=\la-\frac{\mathcal{F}_{11}}{\la}-\frac{\mathcal{F}_{12}}{2\la^2}-\frac{\mathcal{F}_{13}}{3\la^3}-
\frac{\mathcal{F}_{14}}{4\la^4}-\cdots, \label{in}
\end{equation}
where $\mathcal{F}_{1n}$ are polynomials of $v_2,v_3,\cdots,v_{n+1}$ and in fact
\[h_n \equiv \frac{\mathcal{F}_{1n}}{n}=res_{p=\infty} \frac{\lambda^n}{n}dp \]
are  the conserved densities for the dKP hierarchy \eqref{kpp}.
In \cite{bm,Car2}, it's proved that dKP hierarchy \eqref{kpp} is equivalent
to the dispersionless Hirota equation
\begin{equation}
D(\la)S(\la^{'}) =-log  \frac{p(\la)-p(\la^{'})}{\la}, \label{hr}
\end{equation}
where $D(\la)$ is the operator
$\sum_{n=1}^{\infty} \frac{1}{n\la^n}\frac{\pa}{\pa t_n}$. \\
\indent Next, we consider the symmetry constraint \cite{bk, bk1}
\begin{equation}
\mathcal{F}_x=\sum_{i=1}^{n} c_i(S_i-\tilde S_i), \label{st}
\end{equation}
where $S_i=S(\la_i)$ and $\tilde S_i=S(\tilde \la_i)$, $\la_i$, $\tilde \la_i$ are some sets of points, and $c_i$ are arbitrary
constants. Notice that from \eq{in} we know
\[D(\la) \mathcal{F}_x=\la-p.\]
On the other hand, by \eq{hr} and \eq{st}, we also have
\begin{eqnarray*}
D(\la)\mathcal{F}_x &=& D(\la) \sum_{i=1}^{N}c_i(S_i-\tilde S_i)=\sum_{i=1}^{N}c_i(D(\la)S_i-D(\la)\tilde S_i) \\
 &=& -\sum_{i=1}^{N}c_i \log \frac{p-p_i}{p-\tilde p_i},
\end{eqnarray*}
where $p=p(\la), p_i=p(\la_i)$ and $\tilde p_i=p(\tilde \la_i).$ \\
\indent Hence we obtain the non-algebraic reduction("waterbag" model)\cite{bk, gt} of dKP hierarchy
\begin{equation}
\la=p-\sum_{i=1}^{N}c_i \log \frac{p-p_i}{p-\tilde p_i}=p+\sum_{s=1}^{\infty} v_{s+1} p^{-s},
 \label{sa}
\end{equation}
where
\begin{equation}
v_{s+1}=\sum_{i=1}^{N} \frac{c_i(p_i^{s}-\tilde p_i^{s})}{s}.\label{10}
\end{equation}
\indent One remarks that in the limit $\tilde \la_i =\la_i +\epsilon_i$, $\epsilon_i \to 0$, keeping $c_i \epsilon_i=d_i$ to be constant, the Sato function \eq{sa} reproduces the Zakharov's reduction \cite{Za}
\begin{equation}
\la=p+\frac{d_1}{p-\tilde p_1}+\frac{d_2}{p-\tilde p_2}+\cdots+\frac{d_N}{p-\tilde p_N}, \label{za}
\end{equation}
which is the algebraic reduction of dKP hierarchy.\\
\indent From \eqref{sa}, we have
\[B_2(p)=\frac{1}{2}p^2+\sum_{i=1}^{N} c_i(p_i-\tilde p_i).\]
So ($t_2=y$)
\[\pa_y \la=\{\la, \sum_{i=1}^{N} c_i(p_i-\tilde p_i)\}.\]
This will imply
\begin{eqnarray}
\pa_y p_\la &=& \pa_x [\frac{1}{2} p_\la^2 +\sum_{i=1}^{N} c_i(p_i-\tilde p_i)] \label{su} \\
\pa_y \tilde p_\la &=& \pa_x [\frac{1}{2} \tilde p_\la^2 +\sum_{i=1}^{N} c_i(p_i-\tilde p_i)]. \no
\end{eqnarray}
\indent For simplicity, in this paper we only consider the case $N=1$, i.e., \\
\begin{eqnarray}
\pa_y p_1 &=& \pa_x [\frac{1}{2} p_1^2 +c_1(p_1-\tilde p_1)] \label{kp1} \\
\pa_y \tilde p_1 &=& \pa_x [\frac{1}{2} \tilde p_1^2 + c_1(p_1-\tilde p_1)], \no
\end{eqnarray}
and the Lax operator  \eqref{sa} is truncated to become
\begin{equation}
\lambda=p-c_1 \log \frac{p-p_1}{p-\tilde p_1}. \label{lax}
\end{equation}
The equation \eq{kp1} can also be written as the Hamiltonian system
\begin{equation}
 \left[\begin{array}{c}
p_{1y} \\
\\
 \tilde p_{1y}
\end{array} \right]
=\left[\begin{array}{cc}
\frac{1}{c_1} & 0 \\
\\
0 & -\frac{1}{c_1}
\end{array} \right] \pa_{x}
\left[\begin{array}{c}
\frac{\delta H_3}{\delta  p_1} \\
\\
\frac{\delta H_3}{\delta  \tilde p_1}
\end{array} \right],  \no
\end{equation}
where $\delta$ is the variation derivative and
\[H_3=\frac{1}{3} \int dx  [c_1(p_1^3-\tilde p_1^3)+3c_1^2(p_1-\tilde p_1)^2] \]
A bi-Hamiltonian structure is defined as (for the case of dKP)
 \begin{equation}
 \left[\begin{array}{c}
p_{1y} \\
\\
 \tilde p_{1y}
\end{array} \right]
=\mathbb{J}_1
\left[\begin{array}{c}
\frac{\delta H_3}{\delta  p_1} \\
\\
\frac{\delta H_3}{\delta  \tilde p_1}
\end{array} \right]  =\mathbb{J}_2
\left[\begin{array}{c}
\frac{\delta H}{\delta  p_1} \\
\\
\frac{\delta H}{\delta  \tilde p_1}
\end{array} \right], \no
\end{equation}
where
\[\mathbb{J}_1=\left[\begin{array}{cc}
\frac{1}{c_1} & 0 \\
\\
0 & -\frac{1}{c_1}
\end{array} \right] \pa_{x} \]
and $H$ is some Hamiltonian
such that $\mathbb{J}_2$ is also a Hamiltonian operator (Jacobi identity), which is
compatible with $\mathbb{J}_1$, i.e., $\mathbb{J}_1+c \mathbb{J}_2$
is also a Hamiltonian one for any complex number $c$ \cite{du1,du2,Li,Ma}. We hope to find
 $\mathbb{J}_2$ and
the related Hamiltonian $H$.\\
\indent Besides, from \eqref{10} and \eqref{lax} we know that $v_2=v=p_1-\tilde p_1$. Then by
the theory of symmetry constraints of KP \cite{Ch,KS,KST,SS} hierarchy one can consider the dispersive
corrections for the waterbag model \eqref{lax}.  \\
\indent  This paper is organized as follows. In the next section, we construct the
third-order bi-Hamiltonian structure. Section 3 is devoted to establishing first-order
bi-Hamiotonian structure using WDVV equation in topological field theory.
In section 4, we discuss the dispersive corrections. In the final section, one
discusses some problems to be investigated.
\section{Third-Order bi-Hamiltonian Structure}
In this section, we investigate the bi-Hamiltonian structure of the two component case \eqref{kp1}.
To find the bi-Hamiltonian structure, one can  introduce the coordinates
\[ u=p_1 +\tilde p_1, \qquad  v=p_1 -\tilde p_1  \]
to rewrite equation \eqref{kp1} as
\begin{eqnarray}
\left(
\begin{array}{c}
u \\ v
\end{array} \right)_y &=& \left(\begin{array}{c}
\frac{u^2+v^2}{4}+2c_1 v\\ \frac{uv}{2}
\end{array} \right)_x  \label{gg} \\
&=& \frac{1}{c_1} \left( \begin{array}{cc}
0& D_x\\
D_x&0
\end{array}  \right)
\left(\begin{array}{c}
\frac{\delta H_3}{\delta u} \\ \frac{\delta H_3}{\delta v}
\end{array} \right), \no
\end{eqnarray}
where
\begin{eqnarray*}
H_3 &=& \frac{1}{3} \int dx  [c_1(p_1^3-\tilde p_1^3)+3c_1^2(p_1-\tilde p_1)^2] \\
&=&\frac{1}{3} \int dx  [c_1\frac{3u^2 v+ v^3}{4}+3c_1^2v^2]=\int dx  h_3
\end{eqnarray*}
and $\delta$ is the variational derivative. One can observe the conserved density
\[h_3=\frac{1}{3} [c_1\frac{3u^2 v+ v^3}{4}+3c_1^2v^2] \]
has the separable property
\[\frac{\partial^2 h_3}{\partial u^2} / \frac{\partial^2 h_3}{\partial v^2}=
\frac{c_1 v}{2} / (\frac{c_1 v}{2}+2 c_1^2) = \frac{1}{\mu(v)}, \]
where
\[\mu(v)=\frac{v+4c_1}{v}=1+\frac{4c_1}{v}. \]
Hence we can identify the equation \eqref{gg} as the generalized gas dynamic Hamiltonian
system or separable Hamiltonian system \cite{ON,Sh}. Therefore, according to the separable
Hamiltonian theory in \cite{an, ON}, we know that the third-order Hamiltonian operator
\begin{equation*}
J_2= D_x U_x^{-1} D_x U_x^{-1} \sigma D_x,
\end{equation*}
where
\[U_x=\left(\begin{array}{cc}
u_x&\mu(v)v_x \\
v_x&u_x \end{array} \right) \quad  \mbox{and} \quad \sigma=\left( \begin{array}{cc}
0&1 \\
1&0\end{array} \right), \]
or
\[ U_x^{-1}=\frac{1}{u_x^2-\mu(v) v_x^2} \left(\begin{array}{cc}
u_x & -\mu(v)v_x\\
-v_x & u_x\end{array}\right), \]
is compatible with the first order Hamiltonian operator
\[ J_1= \left(\begin{array}{cc}
0&D_x \\ D_x&0\end{array} \right)= \sigma D_x. \]
So we can write the equation \eqref{gg} as the bi-Hamiltonian structure
\begin{eqnarray*}
\left(
\begin{array}{c}
u \\ v
\end{array} \right)_y &=& \frac{1}{c_1} J_1
\left(\begin{array}{c}
\frac{\delta H_3}{\delta u} \\ \frac{\delta H_3}{\delta v}
\end{array} \right)= \frac{1}{c_1} J_2 \left(\begin{array}{c}
\frac{\delta H_5}{3 \delta u} \\ \frac{\delta H_5}{3\delta v}
\end{array} \right) \\
&=&\frac{1}{c_1} D_x \left(\begin{array}{c}
\frac{h_5, uuv}{3 } \\ \frac{h_5, vvu}{3}
\end{array} \right),
\end{eqnarray*}
where
\begin{eqnarray*}
H_5 &=& \int  h_5 dx= \frac{1}{5} \int res_{p=\infty} (\lambda^5dp) dx \\
 &=& \frac{1}{5} \int \{ c_1(p_1^5-\tilde p_1^5)+ 10 c_1^3(p_1-\tilde p_1)^3+
 \frac{20}{3} c_1^2(p_1-\tilde p_1)(p_1^3-\tilde p_1^3)  \\
 &+&\frac{5}{2} c_1^2(p_1^2-\tilde p_1^2)^2 \} dx  \\
&=& \frac{1}{5} \int  \{ \frac{c_1}{16}( 5u^4 v+10 u^2 v^3 +v^5)+10c_1^3 v^3+\frac{15}{2}c_1^2
u^2v^2   \\
&+& \frac{5}{3}c_1^2 v^4 \} dx .
\end{eqnarray*}
\indent Next, one will want to find all the conserved density $F(u,v)$ of equation \eqref{gg}.
It's not difficult to see that if $\int F(u,v) dx$ is a conserved quantity  if and only if
that $F(u,v)$ satisfies the wave equation
\begin{equation}
F_{uu}=\frac{F_{vv}}{\mu(v)} \label{co}
\end{equation}
The wave equation \eqref{co} leads to two fundamental hierarchies of conserved densities \cite{ON, Sh}
\begin{eqnarray}
F_N &=& \sum_{n=0}^{[\frac{N}{2}]} \frac{u^{N-2n}}{(N-2n)!}(\partial_v^{-2} \mu(v))^n \cdot v \no \\
\tilde F_N &=&  \sum_{n=0}^{[\frac{N+1}{2}]} \frac{u^{N+1-2n}}{(N+1-2n)!}(\partial_v^{-2} \mu(v))^n
\cdot 1. \label{sol}
\end{eqnarray}
Here $\partial_v^{-1}=\int_0^v dv$ and $\partial_v^{-1}$ acts on all the factors standing
to the right of it. For example,
\[(\partial_v^{-2} \mu(v))^2 \cdot v= \int_0^v dv \int_0^v [\mu(v) \int_0^v dv
\int_0^v v\mu(v) dv]dv.\]
\indent For reference, we list the first few members of each sequence in the generalized gas dynamic
system \eqref{gg}: ($\mu(v)=1+\frac{4c_1}{v}$)
\begin{eqnarray*}
F_0&=& v, \qquad  F_1=uv  \\
F_2 &=& \frac{1}{2}u^2 v+\frac{1}{6}v^3+2c_1v^2  \\
F_3 &=& \frac{1}{6}u^3 v+ u(\frac{1}{6}v^3+2c_1v^2) \\
F_4 &=& \frac{1}{24} u^4 v+\frac{u^2}{2}(\frac{1}{6} v^3+2c_1v^2)+(\frac{1}{120}v^5+\frac{2}{9}
c_1 v^4+\frac{4}{3}c_1^2v^3) \\
\mbox{and} \\
\tilde F_0 &=& u , \qquad \tilde F_1=\frac{1}{2}u^2+\frac{1}{2}v^2+4c_1(v \ln v-v) \\
\tilde F_2 &=& \frac{1}{6} u^3+u[\frac{1}{2}v^2+4c_1(v \ln v-v)] \\
\tilde F_3 &=& \frac{1}{24} u^4+ \frac{1}{2}u^2[\frac{1}{2}v^2+4c_1(v \ln v-v)] +
\frac{1}{24}v^4+ \frac{2}{3}c_1v^3 \ln v-\frac{8}{9}c_1v^3\\
&+& 8c_1^2v^2 \ln v -20c_1^2 v^2 \\
\tilde F_4 &=& \frac{1}{120} u^5+ \frac{1}{6}u^3[ \frac{1}{2}v^2+4c_1(v \ln v-v) ]
+u[\frac{1}{24}v^4+ \frac{2}{3}c_1v^3 \ln v-\frac{8}{9}c_1v^3  \\
&+& 8c_1^2v^2 \ln v -20c_1^2 v^2]
\end{eqnarray*}
In fact, one can see that
\begin{equation}
F_{N-1}=\frac{2^{N-1}}{c_1 (N-1)!} h_N=\frac{2^{N-1}}{c_1 N!}res_{p=\infty}(\lambda^N dp),\quad
 \mbox{$N\geq 1$}  \label{ser}
\end{equation}
and from \eqref{sol} we also have (for $N \geq 1$)
\begin{eqnarray}
\frac{\partial F_N}{\partial u} &=& F_{N-1}  \no \\
\frac{\partial \tilde F_N}{\partial u} &=& \tilde  F_{N-1} \label{hie}
\end{eqnarray}
Moreover, we notice that  the recursion operator
\begin{equation}
\hat R=J_2 J_1^{-1}= D_x U_x^{-1} D_x U_x^{-1} \label{re}
\end{equation}
is the square of a simpler first-order recursion operator
\[ R=D_x U_x^{-1}.\]
Then we can easily check that , using \eqref{co} and \eqref{hie},
\begin{eqnarray*}
R^{-1} \sigma D \left(\begin{array}{c}
\frac{\partial F_N}{\partial u}\\ \frac{\partial F_N}{\partial v}
\end{array} \right) &=& R^{-1} \left(\begin{array}{c}
\frac{\partial F_N}{\partial v}\\ \frac{\partial F_N}{\partial u}
\end{array} \right)_x =U_x  \left(\begin{array}{c}
\frac{\partial F_N}{\partial v}\\  \frac{\partial F_N}{\partial u}
\end{array} \right) \\
&=&\left(\begin{array}{c}
\frac{\partial F_{N+1}}{\partial v}\\  \frac{\partial F_{N+1}}{\partial u}
\end{array} \right)_x = \sigma D \left(\begin{array}{c}
\frac{\partial F_{N+1}}{\partial u}\\  \frac{\partial F_{N+1}}{\partial v}
\end{array} \right),
\end{eqnarray*}
and similarly for $\tilde F_N$. Hence the dKP hierarchy of \eqref{gg} can be obtained
using the recursion operator, i.e.,
\[\left(\begin{array}{c}
u \\ v  \end{array} \right)_{t_n} = (R^{-1})^{n-2}  \frac{\sigma}{c_1}D
 \left(\begin{array}{c}
\frac{\partial h_3}{\partial u}\\ \frac{\partial h_3}{\partial v}
\end{array} \right), \mbox{$n \geq 2$}. \]
However, the Lax representation of hierarchy generated by $\tilde F_N$ is not found. \\
\indent Also, using the recursion operator \eqref{re}, one can construct a hierarchy of higher
order Hamiltonian densities $\hat F_m $, $m=1,2,3, \cdots ,$ with $m$ indicating the order
of derivatives on which they depend, and the corresponding commuting bi-Hamiltonian system \cite{Pe}:
\begin{equation}
\left(\begin{array}{c}
u \\  v
\end{array} \right)_{\tau_m}= \hat R^m \left(\begin{array}{c}
1\\0 \end{array} \right)= J_1 \left(\begin{array}{c}
\frac{\delta \hat F_m}{\delta u} \\ \frac{\delta \hat F_m}{\delta v}  \end{array} \right)
=J_2 \left(\begin{array}{c}
\frac{\delta \hat F_{m-1}}{\delta u} \\ \frac{\delta \hat F_{m-1}}{\delta v}  \end{array}
 \right), \label{hiee}
\end{equation}
where
\begin{eqnarray*}
\hat F_0 &=& \int xv dx \\
\hat F_1 &=& -\frac{1}{2} \int  \frac{v_x}{u_x^2-\mu(v) v_x^2} dx =
 -\frac{1}{2} \int  \frac{v_x}{u_x^2-(1+\frac{4c_1}{v}) v_x^2} dx.
\end{eqnarray*}
All the flows of \eqref{hiee} will commute with the generalized gas dynamic system \eqref{gg}.
For example, for $n=1$, we have, after a simple calculation,
\begin{eqnarray*}
\left(\begin{array}{c}
u \\  v
\end{array} \right)_{\tau_1} &=& \hat R \left(\begin{array}{c}
1\\0 \end{array} \right)  \\
&=& \left(\begin{array}{c}
\frac{\mu^2v_x^3 v_{xx}+ 3\mu u_x^2v_x v_{xx}+ \mu \mu' v_x^5+\mu' u_x^2 v_x^3-u_x^3 u_{xx}-
3\mu v_x^2 u_x u_{xx}}{(u_x^2-\mu v_x^2)^3}\\ \frac{\mu v_x^3 u_{xx}+ 3 u_x^2v_x u_{xx}-v_{xx}u_x^3-3\mu
v_x^2 u_x v_{xx}-2 \mu' u_x v_x^4}{(u_x^2-\mu v_x^2)^3}
\end{array} \right)_x,
\end{eqnarray*}
where
\[\mu'= \frac{d \mu}{d v}=-\frac{4c_1}{v^2}.\]
\indent Finally, one remarks that there exists a Lagrangian local in the velocity fields
for the equation \eqref{gg}( up to a scaling):
\[\mathcal{L}=\frac{v_x u_t-u_x v_t}{u_x^2-\mu v_x^2}-2v. \]
The local Lagrangian will exist in bi-Hamiltonian structure with a pair of first and third
order Hamiltonian operators \cite{NP, ON}.
\section{Free Energy and Bi-Hamiltonian Structure}
In this section, we investigate the relations between bi-Hamiltonian structure and free energy. Then the compatible first-order Hamiltonian operators can be constructed. \\
\indent Now, we want to find the free energy associated with the dKP hierarchy \eqref{kpp}
for the Lax operator of the form  \eqref{lax}. Suppose we are given two first-order Hamiltonian
operators $\hat J_1$ and $\hat J_2$ ($\partial=\partial_x$)
\bean
\hat J_1 &=&
 \left(
 \ba{cc} 0 & 1 \\
  1 & 0
   \ea
    \right) \pa \overset{def}= \eta_1^{ij} \pa  \no \\
 \hat J_2 &=&
 \left(
  \ba{cc} g^{11}(t) &  g^{12}(t)  \\
    g^{21}(t) &  g^{22}(t)
    \ea
 \right) \pa +\left(
  \ba{cc} \Gamma_1^{11}(t) &  \Gamma_1^{12}(t) \\
   \Gamma_1^{21}(t)& \Gamma_1^{22}(t)
    \ea
 \right)t^1_x+ \left(
  \ba{cc} \Gamma_2^{11}(t) &  \Gamma_2^{12}(t) \\
   \Gamma_2^{21}(t)& \Gamma_2^{22}(t)
    \ea
 \right)t^2_x \notag \\
 & \overset{def}= &g^{ij}(t)\partial +\Ga_{k}^{ij}(t)t^k_x. \notag
\eean
They are both Poisson brackets of hydrodynamic type introduced by Dubrovin and
Novikov \cite{dn, Dn}.
 The bi-Hamiltonian structure means that $\hat J_1$ and $\hat J_2$ have to be compatible, i.e,
 $\hat J=\hat J_1+\alpha \hat J_2$ must also be a Hamiltonian structure for all value of $\alpha$.
 This compatibility
condition implies that, for any $\alpha$, the metric is referred as
flat pencil.  The geometric setting in which to understand flat pencil
( or bi-Hamiltonian structure of hydrodynamic system) is the Frobenius manifold
 \cite{du1,du2, DZ, Du4}.
One way to define such manifolds is to construct
a function $\mathbb{F}(t^1, t^2, \cdots, t^m)$ such that the associated functions,
\[c_{ijk}= \frac{\pa^3 \mathbb{F}}{ \pa t^i \pa t^j \pa t^k},\]
satisfy the following conditions.
\begin{itemize}
\item The matrix $\eta_{ij}=c_{1ij}$ is constant and non-degenerate. This together with the inverse matrix $\eta^{ij}$ are used to raise and lower indices. On such a manifold one may interpret $\eta_{ij}$ as a flat metric.
\item The functions $c_{jk}^{i}=\eta^{ir}c_{rjk}$ define an associative commutative algebra with a unity element. This defines a Frobenius
algebra on each tangent space $T^t \mathcal M$.
\end{itemize}
\indent Equations of associativity give  a system of non-linear PDE
for $\mathbb{F}(t)$
\[\frac{\pa^3 \mathbb{F}(t)}{\pa t^{\alpha} \pa t^{\beta} \pa t^{\lambda}} \eta^{\lambda \mu} \frac{\pa^3 \mathbb{F}(t)}{\pa t^{\mu} \pa t^{\ga} \pa t^{\sigma}}  =
\frac{\pa^3 \mathbb{F}(t)}{\pa t^{\alpha} \pa t^{\ga} \pa t^{\lambda}} \eta^{\lambda \mu} \frac{\pa^3 \mathbb{F}(t)}{\pa t^{\mu} \pa t^{\beta} \pa t^{\sigma}}. \]
These equations constitute the Witten-Dijkgraaf-Verlinde-Verlinde (or WDVV)
equations. On such a manifold one may introduce a second flat metric defined by
\begin{equation}
g^{ij}=\partial^i \partial^j \mathbb{F} + \partial^j \partial^i \mathbb{F}, \label{met}
\end{equation}
where
\[\partial^i=\eta^{i\varsigma}\partial_{t^{\varsigma}}\]
and the contravariant Levi-Civita connection is
\begin{equation}
\Gamma_k^{ij}=\partial^i \partial^j\partial_{t^k} \mathbb{F} , \label{con}
\end{equation}
This metric, together with the original
metric $\eta^{ij}$, define a flat pencil (i.e, $\eta^{ij}+\alpha g^{ij}$ is flat for any
 value of $\alpha$). Thus, one automatically obtains a bi-Hamiltonian structure from a
 Frobenius  manifold $\mathcal M$. The corresponding Hamiltonian densities are defined
  recursively by the formula
\bean
\frac{\pa^2 \psi_{\alpha}^{(n)}}{\pa t^i \pa t^j}=c_{ij}^k \frac{\pa \psi_{\alpha}^{(n-1)}}
{\pa t^k}, \label{rec}
\eean
where $n \geq 1, \alpha =1,2, \cdots, m,$ and $ \psi_{\alpha}^{0}=\eta_{\alpha
\epsilon}t^{\ep}$. The integrability conditions for this systems are automatically
satisfied when the $c_{ij}^k$ are defined as above.\\
\indent For the waterbag hierarchy \eqref{kpp} \eqref{lax}, it is obvious that
\[t^1=u=\psi_2^{(o)},t^2=v=\psi_{1}^{(0)}\]
and those  $c_{ij}^k$ can be determined by \eqref{rec}
\[ \frac{\pa^2 \psi_{1}^{(n)}}{\pa t^i \pa t^j}=c_{ij}^k \frac{\pa \psi_{1}^{(n-1)}}{\pa t^k}, \]
where, using \eqref{ser},
\[\psi_1^{(n)}=F_{n}=\frac{2^n}{c_1 (n+1)!}res_{p=\infty}(\lambda^{n+1} dp), \quad n \geq 0. \]
Simple calculations can get
\bean
c^1_{11}&=& 1,\quad c^1_{12}=c^1_{21}=0,\quad c^1_{22}=1+ \frac{4c_1}{t^2}=\mu(v), \no \\
c^2_{11}&=& c^2_{22}=0, \quad c^2_{21}=c^2_{12}=1, \label{fr}
\eean
By \eqref{fr}, we can get  immediately free energy
\begin{eqnarray}
\mathbb{F}(t^1,t^2)&=&\frac{1}{2}(t^1)^2t^2+2c_1(t^2)^2\log t^2+ \frac{1}{6} (t^2)^3  \no \\
&+ & quadratic \quad terms. \label{free}
\end{eqnarray}
We notice that the free energy \eqref{free} has no quasi-homogeneity condition and, however,
the free energy associated with the Benney hierarchy is quasi-homogeneous \cite{ct}. \\
\indent After choosing suitable quadratic terms, then from the free energy \eqref{free},
using \eqref{met} and \eqref{con}, one can
construct $\hat J_2$ as follows:
\begin{eqnarray*}
\hat  J_2 &=&
 \left(\ba{cc} 2t^2 &  2t^1  \\
    2t^1 &  2t^2
    \ea
 \right) \pa +\left(
  \ba{cc} 0 &  1 \\
  1& 0
    \ea
 \right)t^1_x+ \left(
 \ba{cc} 1 &  0 \\
 0 & 1
    \ea
 \right)t^2_x  \\
&=&\left(\ba{cc} t^2  \partial_x + \partial_x  t^2 & t^1
\partial_x+\partial_x t^1 \\
  t^1 \partial_x+\partial_x t^1    &   t^2 \partial_x+\partial_x t^2
    \ea
 \right).
 \end{eqnarray*}
Here we remove the non-analytic part of $g^{ij}$, i.e., $\ln t^2$. Also,
one can verify directly that $\hat J_2$ is a Hamiltonian operator and is
compatible with $\hat J_1$. We notice that the constant $c_1$ won't appear in $\hat J_2$.  \\
\indent Now, using the recursion operator
\[ \mathbb{\hat R}=\hat J_2 \hat J_1^{-1}= \left(\ba{cc} t^1 +
\partial_x  t^1 \partial_x^{-1} & t^2
+\partial_x t^2\partial_x^{-1} \\
   t^2+\partial_x t^2\partial_x^{-1}  & t^1 + \partial_x  t^1 \partial_x^{-1}
    \ea
 \right), \]
one can construct a hierarchy  by
\begin{equation}
 \left(\begin{array}{c}
t^1 \\  t^2
\end{array} \right)_{\tilde \tau_m} = \mathbb{\hat R}^m \left(\begin{array}{c}
\frac{t^1}{6}\\ \frac{t^2}{6} \end{array} \right)_x, \quad m\geq 1.  \label{hi}
\end{equation}
For example, for $m=1$ and $m=2$, a simple calculation can yield
\begin{eqnarray*}
\left(\begin{array}{c}
t^1 \\  t^2
\end{array} \right)_{\tilde \tau_1} &=& \mathbb{\hat R} \left(\begin{array}{c}
 \frac{t^1}{6}\\ \frac{t^2}{6} \end{array} \right)_x=  \left(\begin{array}{c}
  \frac{1}{4}[(t^1)^2+ (t^2)^2] \\
  \frac{1}{2}t^1 t^2
\end{array} \right)_x, \\
\left(\begin{array}{c}
t^1 \\  t^2
\end{array} \right)_{\tilde \tau_2}& =& \mathbb{\hat R}^2 \left(\begin{array}{c}
\frac{t^1}{6}\\ \frac{t^2}{6} \end{array} \right)_x= 5 \left(\begin{array}{c}
  \frac{1}{12}(t^1)^3+\frac{1}{4}t^1 (t^2)^2 \\
  \frac{1}{12}(t^2)^3+\frac{1}{4}t^2 (t^1)^2
\end{array} \right)_x,
\end{eqnarray*}
which are slightly different  from the $y$ flow \eqref{gg} and $t_3$(or $t$)
flow of the dKP hierarchy \eqref{kpp}, respectively:
\begin{eqnarray}
\left(\begin{array}{c}
t^1 \\  t^2
\end{array} \right)_t & = &\frac{2}{3c_1}  J_1 \left(\begin{array}{c}
\frac{\delta   H_4}{\delta t^1} \\ \frac{\delta  H_4}{\delta t^2}
\end{array} \right)  \no \\
&=& \left(\begin{array}{c}
  \frac{1}{12}(t^1)^3+\frac{1}{4}t^1 (t^2)^2+2c_1t^1 t^2 \\
  \frac{1}{12}(t^2)^3+\frac{1}{4}t^2 (t^1)^2+c_1 (t^2)^2
\end{array} \right)_x , \label{th}
\end{eqnarray}
where
\[H_4=\int h_4 dx=\frac{1}{4} \int \{c_1[\frac{(t^1)^3 t^2+(t^2)^3 t^1}{2}]+6c_1^2 t^1 (t^2)^2\} dx.\]
 \\
By comparisons between them, one can see that the non-homogeneous terms (or higher-
order $c_1$ terms) of the water-bag hierarchy could be  removed in the hierarchy \eqref{hi}.
In this way, one can say that  the hierarchy \eqref{hi} is perturbed, up to some scalings, by the water-bag hierarchy
with a perturbation parameter $c_1$.\\
{\bf Remark:}According to the Kodama-Gibbons formulation \cite{KG}, the Riemann invariants $\lambda_1,
\lambda_2$  of \eqref{th} are given by,
\begin{eqnarray}
\lambda_1&=&\lambda(u_1)=\frac{t^1+\sqrt{(t^2)^2+4 c_1 t^2}}{2}-c_1
\ln \frac{\sqrt{(t^2)^2+4 c_1 t^2}-t^2}{\sqrt{(t^2)^2+4 c_1 t^2}+t^2} \no \\
\lambda_2&=&\lambda(u_2)=\frac{t^1-\sqrt{(t^2)^2+4 c_1 t^2}}{2}+c_1
\ln \frac{\sqrt{(t^2)^2+4 c_1 t^2}-t^2}{\sqrt{(t^2)^2+4 c_1 t^2}+t^2}, \label{ma}
\end{eqnarray}
where $u_1$ and $u_2$ are the real roots of
$\frac{d \lambda}{d p} \vert_{p=u_1, u_2}
=0$, i.e.,
\[u_1=\frac{t^1+\sqrt{(t^2)^2+4 c_1 t^2}}{2}, \quad u_2=\frac{t^1-\sqrt{(t^2)^2+4 c_1 t^2}}{2},\]
and the characteristic speeds are
\[\hat v_1=\frac{d \Omega_3(p)}{dp}\vert_{p=u_1}, \quad \hat v_2=\frac{d \Omega_3(p)}{dp}\vert_{p=u_2}.\]
Then the equation \eqref{th} can be written as
\begin{equation}
\left(\begin{array}{c}
  \lambda_1  \\ \lambda_2
\end{array} \right)_t=\left(\begin{array}{cc}
  \hat v_1&0 \\
  0& \hat v_2
\end{array} \right) \left(\begin{array}{c}
  \lambda_1  \\ \lambda_2
\end{array} \right)_x. \label{riee}
\end{equation}
Also, a simple calculation shows that, using \eqref{ma},  the flat metric $(ds)^2= dt^1 dt^2$
becomes,  in Riemann's invariants,
\[(ds)^2=\eta_{11}(t)(d\lambda_1)^2+ \eta_{22}(t)(d \lambda_2)^2,\]
where
\begin{eqnarray*}
\eta_{11}(t)&=&res_{p_1} \frac{(dp)^2}{d\lambda}=\frac{1}{\frac{d^2 \lambda}{d p^2}
\vert_{p=u_1}}= \frac{t^2}{\sqrt{(t^2)^2+4 t^2 c_1}} \\
\eta_{22}(t)&=&res_{p_2} \frac{(dp)^2}{d\lambda}=\frac{1}{\frac{d^2 \lambda}{d p^2}
\vert_{p=u_2}}= -\frac{t^2}{\sqrt{(t^2)^2+4 t^2 c_1}}.
\end{eqnarray*}
Since it's known that waterbag reduction \eqref{lax} is not scaling invariant \cite{gt},
we can verify that the metric
\begin{eqnarray}
(d \tilde s)^2&=&\frac{\eta_{11}}{u_1}(d\lambda_1)^2+ \frac{\eta_{22}}{u_2}(d \lambda_2)^2 \no \\
&=& \frac{t^2}{4} (dt^1)^2+ \frac{t^1}{2} dt^1 dt^2+ (\frac{t^2}{4}+c_1) (dt^2)^2 \label{ex1}
\end{eqnarray}
is no more flat \cite{Du4}. Hence from the theory of Darboux-Egrov metric \cite{du1}, one
believes that there is no first-order  bi-Hamiltonian structure for \eqref{th}(or \eqref{gg}). However,
we know that \eqref{riee} is (semi-)Hamiltonian \cite{gt, pa} and it probably will
have a compatible non-local Poisson brackets of hydrodynamic type \cite{Maa, MN, Mo1, Mo2, Mo3},
deserving further investigations. \\
\indent Finally, one notices that the metric \eqref{ex1} can also be obtained using the free energy
\eqref{free}
\[\mathbb{F}(t^1,t^2)=\frac{1}{2}(t^1)^2t^2+2c_1(t^2)^2\log t^2+ \frac{1}{6} (t^2)^3
+  2c_1 (t^1)^2.\]
\section{dispersive corrections}
In this section, one will investigate the dispersive corrections of watebag model from the
theory of symmetry constraints of KP hierarchy. \\
\indent Let's briefly describe the KP hierarchy \cite{Di}. The Lax operator of KP hierarchy is
\[L= \partial_X + \sum_{n=1}^{\infty} V_{n+1} \partial_X^{-n}\]
and the KP hierarchy is determined by the Lax equation ($\partial_n=\frac{\partial}{\partial T_n},
T_1= X$)
\[\partial_n L= [B_n, L], \]
where $B_n=\frac{1}{n} L^n_+$ is the differential part of $L^n$. For example($T_2=Y, T_3=T$)
\begin{eqnarray}
V_{2Y}&=&\frac{V_{2XX}}{2}+V_{3X} \label{e1} \\
V_{3Y}&=& \frac{1}{2}V_{3XX}+V_{4X}+V_2 V_{2X} \label{e2}\\
V_{2T}&=&\frac{1}{3} V_{2XXX}+V_{3XX}+V_{4X}+2V_2 V_{2X}. \label{e3}
\end{eqnarray}
Eliminating $V_3$ and $V_4$, we can obtain the KP equation ($V_2=V$)
\begin{equation}
V_T=\frac{1}{4}V_{XXX}+VV_X+ \partial_X^{-1} V_{YY}, \label{KP}
\end{equation}
which also can be described as the compatibility condition for the eigenfunction $\phi$
\begin{eqnarray}
\phi_Y&=&(\frac{1}{2} \partial_X^2+V)\phi \no \\
\phi_T&=& (\frac{1}{3} \partial_X^3+V\partial+V_3 +V_X)\phi. \label{lin}
\end{eqnarray}
To get dKP equation \eqref{dkp}, one simply takes
$T_n \rightarrow \varepsilon T_n=t_n$ in the KP equation \eqref{KP}
, with
\[\partial_{T_n} \rightarrow \varepsilon \partial_{t_n} \quad \mbox{and}
\quad V(T_n) \rightarrow  v(t_n), \]
to obtain the dKP equation when $\varepsilon \rightarrow 0$. Thus the dispersive
term $\frac{1}{4} V_{XXX}$ is removed. Moreover, letting
\[\phi=\exp\frac{S}{\varepsilon}\]
in \eqref{lin}, we also have the equation \eqref{hj} for $n=2, 3$
\begin{eqnarray*}
S_y &=& \frac{1}{2}S_x^2+v_2 \no \\
S_t &=& \frac{1}{3} S_x^3 +v_2 S_x +v_3
\end{eqnarray*}
when $\varepsilon \rightarrow 0$. The compatibility $ S_{ty}=S_{yt}$ will yield
the dKP equation \eqref{dkp}.  \\
\indent  Since $v=p_1-\tilde p_1$, from the theory of symmetry constraints
of KP hierarchy \cite{KS, KST}, one can assume the constraint
\begin{equation}
V=(\ln \phi_1-\ln \phi_2)_X, \label{cont}
\end{equation}
where $\phi_1$ and  $\phi_2$ are arbitrary eigenfunctions, i.e., they both
satisfy equations \eqref{lin}. We remark that if $ \phi_1=\exp\frac{S_1}{\varepsilon}, \quad
 \phi_2=\exp\frac{S_2}{\varepsilon}$ and $ X \rightarrow \varepsilon X=x$, then
 $V(X,Y,T) \rightarrow v(x,y,t)=p_1-\tilde p_1,$ where
 \[p_1=S_{1x}, \quad \tilde p_1= p_2=S_{2x}\]
when $\varepsilon \rightarrow 0$. \\
\indent Then equations \eqref{lin} become, $i=1,2$,
\begin{eqnarray}
\phi_{iy}&=&\frac{1}{2} \phi_{iXX}+(\ln \phi_1-\ln \phi_2)_X \phi_i \label{co1} \\
\phi_{iT}&=&\frac{1}{3} \phi_{iXXX}+(\ln \phi_1-\ln \phi_2)_X \phi_{iX} \label{co2} \\
&+& [V_3+(\ln \phi_1-\ln \phi_2)_{XX}] \phi_i. \no
\end{eqnarray}
By equations \eqref{e1} and \eqref{co1}, a simple calculation yields
\begin{equation}
V_3=\frac{[(\ln \phi_1)_X]^2-[(\ln \phi_2)_X]^2}{2} \label{qu1}
\end{equation}
and by equations \eqref{e2} and \eqref{co1}, one obtains
\begin{equation}
V_4=\frac{1}{3} \{[(\ln \phi_1)_X]^3-[(\ln \phi_2)_X]^3\}. \label{qu2}
\end{equation}
One notices that the dispersionless limits of \eqref{qu1} and \eqref{qu2}
are $v_3=\frac{p_1^2-\tilde p_1^2}{2}$ and $v_4=\frac{p_1^3-\tilde p_1^3}{3}$
, respectively. Unfortunately, using \eqref{e3} and \eqref{co2}, after a tedious calculation,
one has
\[[(\ln \phi_1)_{XX}]^2-[(\ln \phi_2)_{XX}]^2=0,  \]
which is a contradiction since $\phi_1$ and $\phi_2$ are arbitrary eigenfunctions.
This means that \eqref{co2} is not a higher-order Lie-Backlund symmetry of \eqref{co1}
or the constraint \eqref{cont} is not admissable\cite{KS, SS}. \\
\indent Also, we can consider the following case generalizing \eqref{cont}
\begin{equation}
V=[1-f(\partial_X)]^{-1}[\ln \phi_1-\ln \phi_2]_X \label{ge}
\end{equation}
or
\[V-f(\partial_X)V=[\ln \phi_1-\ln \phi_2]_X, \]
where
\[f(\partial_X)=a_1\partial_X+ a_2 \partial_X^2+\cdots+ a_n\partial_X^n,
\quad \mbox{$a_i$ being constants}.\]
The dispersionless limit of \eqref{ge} is also $v=p_1-\tilde p_1$. Similarly,
from \eqref{e1} and \eqref{e2} and \eqref{co1}, one can get
\begin{eqnarray*}
V_3 &=&\frac{1}{2}[1-f(\partial_X)]^{-1}\{[(\ln \phi_1)_X]^2-[(\ln \phi_2)_X]^2\} \\
V_4 &=&\frac{1}{3}[1-f(\partial_X)]^{-1}\{ [(\ln \phi_1)_X]^3-[(\ln \phi_2)_X]^3 \\
&+& 3 [f(\partial_X)V^2]-3V[f(\partial_X)V] \} .
\end{eqnarray*}
Both the dispersionless limits of $V_3$ and $V_4$  are
 $v_3=\frac{p_1^2-\tilde p_1^2}{2}$ and $v_4=\frac{p_1^3-\tilde p_1^3}{3}$. \\
\indent Now, using \eqref{e3} and \eqref{co2}, a lengthy calculation shows that
\begin{equation}
2V[ f(\partial_X)V_X]-[ f(\partial_X)(V^2)_X]=
[(\ln \phi_1)_{XX}]^2-[(\ln \phi_2)_{XX}]^2, \label{fi}
\end{equation}
which is also an contradiction. Hence in the general case \eqref{ge}, it's
also not admissible. For example, letting $f(\partial_X)=\partial_X$, we have ($V_2=V$)
\begin{eqnarray}
V &=& (1-\partial_x)^{-1}[\ln \phi_1-\ln \phi_2]_X \label{ex} \\
V_3 &=& \frac{1}{2}[1-\partial_X]^{-1}\{[(\ln \phi_1)_X]^2-[(\ln \phi_2)_X]^2\} \no \\
V_4 &=& \frac{1}{3}[1-\partial_X]^{-1}\{ [(\ln \phi_1)_X]^3-[(\ln \phi_2)_X]^3 +3V_X \} \no
\end{eqnarray}
and \eqref{fi}
becomes
\[ -2 V_X^2=[(\ln \phi_1)_{XX}]^2-[(\ln \phi_2)_{XX}]^2 \]
or, using \eqref{ex},
\[(\ln \phi_{1}-\ln \phi_2)_{XX}=\pm (1-\partial_X)\sqrt{\frac{
[(\ln \phi_2)_{XX}]^2-[(\ln \phi_1)_{XX}]^2}{2}}, \]
which puts constraint on $\phi_1$ and $\phi_2$. Therefore \eqref{ex}
is not admissible.
\section{concluding Remarks}
We  have investigated the bi-Hamiltonian structure and dispersive corrections of the waterbag model
for two component. After introducing suitable coordinates, one can identify \eqref{gg} as a
separable Hamiltonian system and thus a third-order bi-Hamiltonian structure is obtained.
Also, using the recursion relation of conserved densities , we can find the free energy associated
with water-bag model in WDVV equation of the topological field theory and thus establish
the first-order bi-Hamiltonian structure. However,the hierarchy constructed
by recursion operator is not the same as dKP hierarchy. Also, one considers the dispersive
corrections from the theory of symmetric constraints of KP theory. Some calculations show that
these dispersive corrections are not admissible. Finally, one remarks that the solutions
 of the waterbag model can be found using the hodograph method in \cite{KG, ko}. \\
\indent  Several   questions remain to be overcome.
Firstly, from the theory of non-local Poisson brackets of hydrodynamic type \cite{Mo1,Mo3},
one believes
the bi-Hamiltonian structure of \eqref{th} (or \eqref{gg})deserves further investigations, especially that
the free energy \eqref{free} is not quasi-homogeneous. Secondly, as we see in section 4,
the integrable dispersive corrections of the waterbag model  are still unknown.
The main difficulty is in the quantization of the Lax operator \eqref{sa}, i.e., $p \rightarrow
\partial_X $. The exact form is not clear and needs further investigations \cite{St}.
Ultimately, we hope to generalize the results in section 2 to the general case, for example,
four-component case. But the computation is more involved. One hopes to address these questions
elsewhere. \\
{\bf Acknowledgments\/} \\
The author is grateful to Professor Maxim V.Pavlov for his stimulating discussions on bi-Hamiltonian
theory. He also thanks  Professor I.A.B. Strachan for  valuable suggestions.
He thanks the National Science Council for support under grant no. NSC 94-2115-M-014-001.

\end{document}